\documentclass[]{article}

\usepackage{graphicx}
\usepackage{amsmath}
\usepackage{amssymb}
\usepackage{hyperref}
\hypersetup{
colorlinks,
citecolor=blue,
filecolor=blue,
linkcolor=blue,
urlcolor=blue}

\newcommand{\be}{\begin{equation}}
\newcommand{\ee}{\end{equation}} 
\newcommand{\bea}{\begin{eqnarray}}
\newcommand{\eea}{\end{eqnarray}}

\begin{document}

\title{Applicability  of Taylor's hypothesis in thermally driven turbulence}

\date{}

\author{
Abhishek Kumar\footnote{Applied Mathematics Research Centre, Coventry University, Coventry CV15FB, UK}~\footnote{abhishek.kir@gmail.com}~~and~~Mahendra K. Verma\footnote{Department of Physics, Indian Institute of Technology Kanpur, Kanpur 208016, India}}
\maketitle




\begin{abstract}
In this paper, we show that in the presence of large-scale circulation (LSC), Taylor's hypothesis can be invoked to deduce the energy spectrum in thermal convection using real space probes, a popular experimental tool. We perform numerical simulation of turbulent convection in a cube and observe that the velocity field follows Kolmogorov's spectrum ($k^{-5/3}$).  We also record the velocity  time series using real space probes near the lateral walls.  The corresponding frequency spectrum exhibits Kolmogorov's   spectrum  ($f^{-5/3}$), thus validating Taylor's hypothesis with the steady LSC playing the role of a mean velocity field.  The aforementioned findings based on real space probes provide valuable inputs for experimental measurements used for studying the spectrum of convective turbulence.
\end{abstract}

\section{Introduction}
\label{sec:intro}

Thermal convection exhibits a wide range of phenomena---instabilities, patterns, chaos, and turbulence, depending on the strength of the buoyancy force. An idealized system called {\em Rayleigh-B\'{e}nard convection} (RBC)~\cite{Chandrasekhar:book,JKB:book,Bodenschatz:ARFM2000,Ahlers:RMP2009,Lohse:ARFM2010} in which a thin layer of fluid is heated from below and cooled from the top captures the aforementioned complexity.  Turbulent convection, a topic of this article, remains largely unsolved despite a century of efforts. In this paper, we  discuss the spectral properties of the velocity { and temperature fields} in RBC.  The two important parameters of RBC are the Rayleigh number, which is defined as the ratio of the buoyancy  and the viscous term, and the Prandtl number, which is the ratio of the kinematic viscosity and thermal diffusivity.

For isotropic hydrodynamic turbulence, Kolmogorov~\cite{Kolmogorov:DANS1941a} showed that one-dimensional energy spectrum $E(k)=C\epsilon^{2/3}k^{-5/3}$, called Kolmogorov spectrum, for the intermediate range of wavenumbers ($k$).  Here $\epsilon$ is the energy flux, and $C$ is the Kolmogorov's constant.  For buoyancy-driven turbulence with stable stratification, Bolgiano~\cite{Bolgiano:JGR1959} and Obukhov~\cite{Obukhov:DANS1959} argued that in the wavenumber band $k < k_B$, $E(k)\sim k^{-11/5}$ for the velocity field and $E_T(k) \sim k^{-7/5}$ for the temperature field, but in the wavenumber band  $k_B < k < k_d$, both velocity and temperature fields  exhibit Kolmogorov's spectrum. Here $k_B, k_d$ are the Bolgiano and Kolmogorov's wavenumber respectively.   The steepening of the velocity spectrum for $k < k_B$ is due to the conversion of the kinetic energy to the potential energy that depletes the energy flux to yield $\Pi(k) \sim k^{-4/5}$ \cite{Bolgiano:JGR1959,Obukhov:DANS1959}.

Procaccia and Zeitak~\cite{Procaccia:PRL1989} and L'vov and Falkovich~\cite{Lvov:PD1992} argued that the aforementioned Bolgiano--Obukhov phenomenology of stably stratified turbulence also applies to RBC.   Recently Kumar {\em et al.}~\cite{Kumar:PRE2014} and Verma {\em et al.}~\cite{Verma:NJP2017} showed that the turbulence phenomenology of RBC differs significantly from that of stably stratified turbulence; in RBC,  the temperature field feeds the kinetic energy, hence the kinetic energy flux is a nondecreasing function of wavenumber, rather than decreasing as $k^{-4/5}$. {  For unit Prandtl number,  numerical simulations of Kumar {\em et al.}~\cite{Kumar:PRE2014} and Verma {\em et al.}~\cite{Verma:NJP2017}  show that the pressure gradient dominates the buoyancy and viscous dissipation, hence turbulent convection has similar physics as three-dimensional hydrodynamic turbulence. In addition, viscous dissipation tends to balance the energy feed by buoyancy.  These effects make the kinetic energy flux a constant in the inertial range that leads to Kolmogorov's spectrum for RBC.} A shell model for RBC~\cite{Kumar:PRE2015} also confirms the above observations, albeit at larger Rayleigh numbers. { It is important to note that the temperature spectrum $E_T(k)$ for turbulent convection exhibits a dual branch.   The upper branch of the spectrum is proportional to $k^{-2}$, while the lower branch does not exhibit a clearcut power law~\cite{Mishra:PRE2010,Kumar:PRE2014,Verma:NJP2017}.  This observation causes doubt on the usage of the temperature field for testing whether Bolgiano--Obukhov ($E_T \sim k^{-7/5}$) or Kolmogorov--Obukhov ($E_T \sim k^{-5/3}$) is applicable for turbulent convection.   This is confounded by the fact that the competing spectral indices for the temperature field, $-5/3$ and $-7/5$ are too close to each other for an easy contrast. }

To probe turbulence in thermal convection, scientists measure and analyze the velocity and temperature fields in experiments. The determination of the energy spectrum $E(k)$ requires complete three-dimensional high-resolution real-space data, which is difficult to record at present. Only a handful RBC experiments captured two-dimensional (2D) high-resolution velocity field using 2D particle image velocimetry~\cite{Sun:PRL2006,Zhou:JFM2008,Kunnen:PRE2008,Zhou:JFM2011}; an approximate energy spectrum is computed from such data under the assumption of homogeneity and isotropy,  which is not  strictly valid in convection. In most experiments, the velocity field, $u_z(t)$, and/or temperature field, $T(t)$, are probed at fixed points in the flow~\cite{Wu:PRL1990,Chilla:INCD1993,Cioni:EPL1995,Zhou:PRL2001,Skrbek:PRE2002,Niemela:NATURE2000,Shang:PRE2001}.

In a fluid moving with a constant velocity ${\bf U}_0$,  Taylor's hypothesis~\cite{Taylor:PRSLA1938} is invoked to relate the frequency power spectrum, $E(f) = |u(f)|^2/2$,  to one-dimensional wavenumber spectrum $E(k)$ using $E(f) = E(k)(2 \pi )/U_0$ since $f = U_0 k/(2 \pi)$. In Appendix~\ref{appA}, we show that in hydrodynamic turbulence with significantly large $U_0$, $E(k) \sim k^{-5/3}$ and $E(f) \sim f^{-5/3}$ in accordance with Taylor's hypothesis. In addition,  $E(k) \approx \tilde{E}(\tilde{f})$ with appropriate scaling,  frequency $f \rightarrow \tilde{f} = f (2\pi)/U_0$ and $E(f) \rightarrow \tilde{E}(\tilde{f}) = E(f) U_0/(2\pi)$ (see Appendix~\ref{appA} for details). However, for homogeneous and isotropic turbulence with ${\bf U}_0=0$, we show that  $E(f) \sim f^{-2}$ since $f \sim \epsilon^{1/3}k^{2/3}$ from Kolmogorov's theory~\cite{Tennekes:book,Landau:Book}.  In a related development, elliptic approximation has been used  to relate spatial and temporal Eulerian two-point correlations  in the absence of mean flow; the above computation retains the effects of sweeping by the large eddies~\cite{He:PRE2006,He:Acta2014,He:ARFM2016,Cholemari:JFM2006}.

Unfortunately, thermal convection in a box does not have a mean velocity ${\bf U}_0$; hence an application of Taylor's hypothesis to convective turbulence has been intensely debated~\cite{Ahlers:RMP2009,Lohse:ARFM2010}.  Lohse and Xia~\cite{Lohse:ARFM2010} argue that velocity in the central region vanishes, while it is close to the root-mean-square (rms) velocity near the sidewalls, hence, as argued by  Lohse and Xia~\cite{Lohse:ARFM2010}, "the condition for the Taylor hypothesis is often not met in turbulent RB convection, and its applicability to the system is at best doubtful."  As argued previously, most experiments, however, measure velocity and/or temperature fields at the select number of probes; hence the conclusive study of the applicability of Taylor's hypothesis is crucial. 

Recently He {\em et al.}~\cite{He:PRE2010} and He and Tong~\cite{He:PRE2011} attempted to verify Taylor's hypothesis in turbulent convection; they used the well-known elliptic approximation~\cite{He:PRE2006,He:Acta2014,He:ARFM2016} that combines the local mean velocity and the random sweeping velocity. First, they computed the temperature correlation function 
\begin{equation}
C_T(r,\tau)=\frac{\langle \delta T (x+r,t+\tau) \delta T(x,t)\rangle_t}{(\sigma_T)_1(\sigma_T)_2},
\end{equation}
where $r$ is the spatial position of the probe, $\tau$ is time separation,  $\delta T$ is the local temperature deviation from the mean and $(\sigma_T)_i$ is its standard deviation at position $i$. They relate the above correlation to equal-time correlation $C_T(r_E,0)$ using
\begin{equation}
C_T(r,\tau) = C_T(r_E,0),
\end{equation}
where $r_E$ is of the following elliptic form
 \begin{equation}
r_E^2 = (r-U\tau)^2 + (V\tau)^2.
\label{eq:rE}
\end{equation}
Here $U$ is  the local mean velocity, and $V$ is associated with a random sweeping velocity.  After this, He {\em et al.}~\cite{He:PRE2010} compute the one-dimensional energy spectrum $E_T(k)$ by taking Fourier transform of $C_T(r_E,0)$, and obtained $E_T(k)  \sim k^{-1.35}$.  This computation, though having performed using the well-known elliptic approximation, does not capture the Kolmogorov-like spectrum for the velocity as reported recently by Kumar {\em et al.}~\cite{Kumar:PRE2014} and Verma {\em et al.}~\cite{Verma:NJP2017}.  The divergence possibly occurs due to the usage of the temperature field that exhibits dual spectrum because of the boundary layer~\cite{Mishra:PRE2010,Kumar:PRE2014,Verma:NJP2017}.  This difficulty necessitates a revisit of  Taylor's hypothesis in turbulent convection.  In this paper, we focus on the numerical study of the velocity field for which the energy spectrum $E(k)$ is quite unambiguous~\cite{Kumar:PRE2014,Verma:NJP2017}.

A lack of clarity in the application of Taylor's hypothesis for convective turbulence is one of the biggest stumbling blocks for understanding convective turbulence, especially for the   spectra of the velocity and temperature fields. Chill\`a {\em et al.}~\cite{Chilla:INCD1993} and Zhou and Xia~\cite{Zhou:PRL2001} measured the time series of the temperature field in convection experiments on water and reported Bolgiano--Obukhov scaling. Wu {\em et al.}~\cite{Wu:PRL1990} also reported Bolgiano--Obukhov scaling from the frequency spectrum of the temperature field for helium gas. Castaing~\cite{Castaing:PRL1990} and  Cioni {\em et al.}~\cite{Cioni:EPL1995} however reported Kolmogorov's scaling for the temperature field in the helium gas and mercury experiments respectively. Shang and Xia~\cite{Shang:PRE2001} and Mashiko {\em et al.}~\cite{Mashiko:PRE2004} reported Bolgiano--Obukhov scaling from the time series of the velocity field of water and mercury respectively. Ashkenazi and Steinberg~\cite{Ashkenazi:PRL1999b} performed an experiment with sulphur hexafluoride ($\mathrm{SF}_6$) gas and reported Bolgiano--Obukhov scaling in the frequency spectra for both temperature and velocity fields. Niemela {\em et al.}~\cite{Niemela:NATURE2000} reported a dual scaling, as predicted by Bolgiano and Obukhov, from the probe measurement of the temperature field for helium gas. Skrbek {\em et al.}~\cite{Skrbek:PRE2002} computed the temperature structure functions in time domain with the cryogenic helium gas as working fluid and obtained  scaling exponents in the Bolgiano regime.  Using the above data, Bershadskii {\em et al.}~\cite{Bershadskii:PRE2004} obtained $E_T(f) \sim f^{-1.37}$  which they relate to the Clusterization and intermittency.  

Apart from the time-domain measurements, space-domain measurements were also carried by the researchers~\cite{Sun:PRL2006,Zhou:JFM2008,Kunnen:PRE2008,Zhou:JFM2011} using 2D particle image velocimetry (PIV). Sun {\em et al.}~\cite{Sun:PRL2006} observed Kolmogorov's scaling in the central region of the cell for water. Kunnen {\em et al.}~\cite{Kunnen:PRE2008} analyzed the scaling of structure function for water and observed Bolgiano--Obukhov scaling.  The scaling of energy spectrum  for convective turbulence has also been studied by creating  density difference in a long vertical tube~\cite{Arakeri:CS2000}.  Pawar and Arakeri~\cite{Pawar:POF2016} created density difference by using the brine in bottom tank and fresh water in the top tank and achieved $\mathrm{Ra}\approx 10^{10}$ with $\mathrm{Pr} \approx 600$ and shows KO  scaling for the velocity field  and BO scaling for the concentration fluctuation. The above results indicate significant uncertainties on the determination of the spectrum of convective turbulence.

Numerical simulations of RBC provide access to complete velocity field, but lower resolution and ideal boundary conditions used in numerical simulations  hinder clearcut determination of $E(k)$. Grossmann and Lohse~\cite{Grossmann:PRL1991} performed the simulation for $\mathrm{Pr}=1$ under Fourier-Weierstrass approximation and reported Kolmogorov's scaling. Based on periodic boundary condition,  Borue and Orszag~\cite{Borue:JSC1997} and \v{S}kandera {\em et al.}~\cite{Skandera:HPCISEG2SBH2009} reported Kolmogorov's scaling for the velocity and temperature fields. Rincon~\cite{Rincon:JFM2006} performed simulation for $\mathrm{Pr}=1$ and $\mathrm{Ra}=10^6$ using a higher order finite-difference scheme.  He employed the SO(3) analysis to treat isotropic and anisotropic projections of the structure function, but his analysis was inconclusive in identifying any definite spectral slope.  For zero and small Prandtl numbers, Mishra and Verma~\cite{Mishra:PRE2010}  showed that  $E(k) \sim k^{-5/3}$ since the buoyancy is essentially concentrated near the low wavenumbers for such flows, similar to that in hydrodynamic turbulence that exhibits $k^{-5/3}$ energy spectrum. 

Verzicco and Camussi~\cite{Verzicco:JFM2003} and Camussi and Verzicco~\cite{Camussi:EJMF2004} performed numerical simulation  in a cylindrical geometry and collected the data from the real space probe. The frequency spectrum from numerical data exhibit Bolgiano--Obukhov scaling. Interestingly, Calzavarini {\em et al.}~\cite{Calzavarini:PRE2002} observed both  Bolgiano--Obukhov and Kolmogorov's spectra in the boundary layer and bulk respectively. Recently, De {\em et al.}~\cite{De:IJHFF2017} also observed similar variations in the velocity field exponent. Kaczorowski and Xia~\cite{Kaczorowski:JFM2013} performed the simulation for $\mathrm{Pr} = 0.7$ and $4.38$ for Rayleigh number ranging from $10^5$ to $10^9$ and reported KO scaling for the longitudinal velocity structure functions, but BO scaling for the temperature structure functions in the centre of the cubical cell. Kerr~\cite{Kerr:JFM1996} performed the simulation for $\mathrm{Pr} \approx 1$ on a $288 \times 288 \times 96$ grid in a cubical box using Chebyshev based pseudospectral method under no-slip boundary conditions; he reported the horizontal spectrum as a function of horizontal wavenumber $k_{\perp} = \sqrt{k_x^2 + k_y^2}$ and observed Kolmogorov's spectrum.     Recently Nath {\em et al.}~\cite{Nath:up2016} showed that the convective turbulence is weakly anisotropic.

{  The aforementioned works cast doubt on which type of experiments on turbulent convection are suitable for probing the energy and entropy (of temperature field) spectra of the flow.  These spectra carry a signature that  tells us which of the two scaling, Kolmogorov--Obukhov  or Bolgiano--Obukhov, is valid for turbulent convection.  Since Taylor's hypothesis is questionable for turbulent convection, an experimentalist may not opt for measurements using real space probes, and choose 3D or 2D PIV.  However, the resolutions of present-day PIV setups  are not very high, hence they may not yield the desired spectrum. In addition, PIV experiments are much more expensive than probe measurements.}

{  Considering the above issues, we attempt to figure out  regimes and geometries of turbulent convection for which Taylor's hypothesis may be applicable.  In the  present paper, we show that Taylor's hypothesis is applicable to turbulent convection {\em only} when a {\em steady} large-scale circulation (LSC) is present in the flow. For the aforementioned purpose, we performed simulation in a cube for  Prandtl number, $\mathrm{Pr} = 1$,  and Rayleigh number  $\mathrm{Ra} = 10^8$.  For these parameters, we observe  a steady  large-scale circulation~\cite{Niemela:JFM2001,Sreenivasan:PRE2002,Brown:PRL2005,Xi:PRE2007,Mishra:JFM2011}. For the velocity field, we compute the wavenumber spectrum, as well as the frequency spectrum from the time series measured by a set of real-space probes.  We show that both these spectra follow Kolmogorov's $k^{-5/3}$ spectrum. Thus, we show that  Taylor's hypothesis is valid for such system due to the {\em local} constant velocity near the lateral walls. Note that we are only considering the local mean velocity $U$, not the random sweeping velocity $V$ [see Eq.~(\ref{eq:rE})] in the present analysis.}

{ Note that thermal convection in a cylinder exhibits azimuthal reorientations and reversals of LSC~\cite{Niemela:JFM2001,Sreenivasan:PRE2002,Brown:PRL2005,Xi:PRE2007,Mishra:JFM2011}. As described earlier, these movements would make Taylor's hypothesis inapplicable for the cylindrical geometry.  Hence, we believe that for probing the energy spectrum in turbulent convection, a rectangular geometry is a better candidate than a cylinder.  However, recent large-eddy numerical simulations~\cite{Foroozani:PRE2017} of thermal convection in  a cube for $\mathrm{Pr} = 0.7$, and $\mathrm{Ra} = 10^8$ exhibits flow reversals.  Note that, Vasiliev~{\em et al.}~\cite{Vasiliev:IJHMT2016} observed random reorientations of LSC in a cubic cell; they also studied the sensitivity of LSC on experimental design. Hence, we need to carry out further analysis to test whether Taylor's hypothesis will be applicable in a cube in which LSC exhibits flow reversals.}

The outline of the paper is as follows. In Sec.~\ref{sec:gov} we set up our governing equations. In Sec.~\ref{sec:sim} we explain our simulation methods and discuss the results of our numerical simulations   in Sec.~\ref{sec:results}. We conclude in Sec.~\ref{sec:conc}.

\section{Governing equations}
\label{sec:gov}

The dynamical equations that describe RBC under Boussinesq approximation are 
\begin{eqnarray}
\frac{\partial \bf u}{\partial t} + (\bf u \cdot \nabla) \bf u & = & - \frac{1}{\rho_0}\nabla p+ \alpha g T \hat{z} + \nu\nabla^2 \bf u , \label{eq:u} \\
\frac{\partial T}{\partial t} + ({\bf u \cdot \nabla}) T & = &  \kappa \nabla^2 T, \label{eq:T} \\
\nabla \cdot \bf u & = & 0 \label{eq:inc}, 
\end{eqnarray}
where ${\bf u}$ and $T$ are the velocity and temperature fields respectively, and $\hat{z}$ is the buoyancy direction. Here $\alpha$ is the thermal expansion coefficient, $g$ is the acceleration due to gravity, and $p$ is the pressure field, and $\rho_0$, $\nu$, $\kappa$ are the fluid's mean density, kinetic viscosity, and thermal diffusivity respectively. 

It is convenient to work with nondimensionalized equations. We nondimensionalize Eqs.~(\ref{eq:u})-(\ref{eq:inc}) using $d$ as the length scale, the large-scale velocity $(\alpha g \Delta d)^{1/2}$ as velocity scale, and $\Delta$ as the temperature scale, where $\Delta$ and $d$ are  the  temperature difference and the distance between the plates respectively.  The eddy turnover time is the time scale of our simulation.  The nondimensional equations are 
 \begin{eqnarray}
\frac{\partial \bf u}{\partial t} + (\bf u \cdot \nabla) \bf u & = & - \nabla p+ T \hat{z} + \sqrt{\frac{\mathrm{Pr}}{\mathrm{Ra}}} \nabla^2 \bf u , \label{eq:non_u} \\
\frac{\partial T}{\partial t} + ({\bf u \cdot \nabla}) T & = &  \frac{1}{\sqrt{\mathrm{RaPr}}} \nabla^2 T, \label{eq:non_T} \\
\nabla \cdot \bf u & = & 0 \label{eq:non_inc}.
\end{eqnarray}
The two nondimensional control parameters are the Prandtl number $\mathrm{Pr}=\nu/\kappa$ and the Rayleigh number $\mathrm{Ra} = \alpha g \Delta d^3/(\nu \kappa)$. 

In this paper we solve the above equations numerically and study the energy spectrum for the velocity field in wavenumber space:
\begin{eqnarray}
E(k) &=& \sum_{k -1 < k^{\prime} \leq k} \frac{1}{2} |\hat{\bf u}({\bf k^\prime})|^2. \label{eq:KE_spectrum} 
\end{eqnarray}
Then we compare $E(k)$  with the frequency spectrum computed using the time series measured by the real space probes. { The real space probes are used to measure the velocity or temperature fields at particular locations in the real space, as exhibited in Fig.~\ref{fig:probe} (see Sec.~\ref{sec:sim} for details).} For better averaging, we employ multiple number of probes in the neighborhood and take the average of the measured signal as
\begin{equation}
u_i(t) = \frac{1}{n} \sum_k u_{i,l}(t),
\label{eq:u_i}
\end{equation}
where $i$ stands for the velocity component $(i=x,y,z)$, and $l$ stands for the probe index.   We compute the frequency spectrum $E(f)$ of the velocity field as
\begin{equation}
E(f) = \frac{1}{2} \left( |\hat{u}_x(f)|^2 + |\hat{u}_y(f)|^2 +  |\hat{u}_z(f)|^2 \right),
\label{eq:Ef}
\end{equation}
 where $\hat{u}_i$ is the Fourier transform of the $i$-th component of  the velocity field. Induction of more probes is to decrease the fluctuations in $E(f)$ since $\sigma_n =\sigma/\sqrt{n}$, where $\sigma$ and $\sigma_n$ are the standard deviations with single  probe and $n$ probes respectively.  Further, to reduce noise in the frequency spectrum, we perform time-windowed averaging~\cite{Madhow:book}. We break the velocity time-series data of a real space probe into 8 windows and then compute the frequency spectrum of each window using Eq.~(\ref{eq:Ef}). We report the frequency spectrum averaged over these windows for a real space probe. { Note that, we also compute the temperature field $T$ at various probe locations, similar to the velocity field,  and compute the entropy spectrum  $E_T(f)$ in the frequency space and $E_T(k)$ in the wavenumber space.}
 
Another important quantity of RBC is the kinetic energy flux $\Pi(k_0)$,  which is defined as the kinetic energy leaving a wavenumber sphere of radius $k_0$ due to nonlinear interactions.  The kinetic energy flux is computed  using the formula~\cite{Dar:PD2001,Verma:PR2004}
\begin{equation}
\Pi(k_0)  =  \sum_{k > k_0} \sum_{p\leq k_0} \Im([{\bf k \cdot u(k-p)}]  [{\bf u^*(k) \cdot u(p)}]). \label{eq:ke_flux}
\end{equation}
In Kolmogorov's theory of turbulence, $\Pi(k_0) $ is a constant in the inertial range, and it is equal to the viscous dissipation rate.

\section{Simulation methods}
\label{sec:sim}
Equations~(\ref{eq:non_u})-(\ref{eq:non_inc})  are solved in a closed cubical box of unit dimension using an open-source finite-volume code OpenFOAM~\cite{OpenFOAM}. We employ the no-slip boundary condition for the velocity field at all the walls, conducting  boundary conditions for the temperature field at the horizontal wall, and insulating boundary condition at the vertical wall. Gaussian finite volume integration is used for the computation of derivative terms ($\nabla p$, convective, and Laplacian). Gaussian integration is based on a sum of the values of a function on the cell faces; these values are interpolated from the cell centres to the nodes. These  data at the cell centres are interpolated using linear interpolation. For time stepping we use second order Crank-Nicolson scheme. 

\begin{figure}[htbp]
\begin{center}
\includegraphics[scale = 0.8]{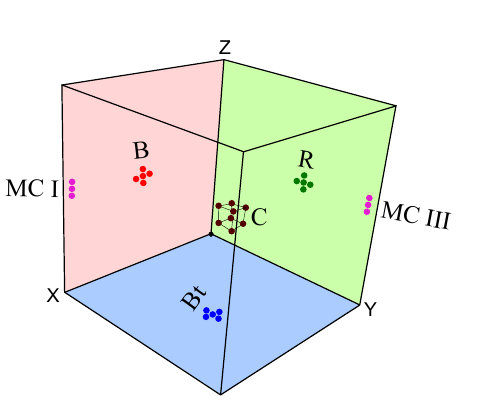}
\end{center}
\caption{The real space probe locations in the three-dimensional cubical box of our simulation. {The probe locations are labeled as the back (B), right (R), bottom (Bt), middle corners (MC-I, MC-III), and centre (C) respectively.} There are $5$ probes near the wall centres,  $3$ probes in the middle corners, and $9$ probes at the centre \& vertices of a small cube placed at the centre of the cube.}
\label{fig:probe}
\end{figure}

We perform  simulation for $\mathrm{Pr}=1$ (close to that of air) and $\mathrm{Ra}=10^8$. The grid resolution of our simulation is $256^3$ in a nonuniform mesh with a higher grid concentration near the boundaries in order to resolve the boundary layer. The Reynolds number $\mathrm{Re}$ for this run is approximately $1634$.  An important response parameter for the convective turbulence is the Nusselt number $\mathrm{Nu}$, which is the ratio of the total (convective plus conductive) heat flux and the  conductive heat flux.  For the aforementioned simulation,  $\mathrm{Nu} \approx 34.4$. We employ a constant $\Delta t=10^{-3}$ for which the Courant number is less than unity.  Here $t=1$ of our simulation corresponds to $d/\sqrt{\alpha g \Delta d}$.  Note that the aforementioned constant $\Delta t$ helps us to compute the Fourier transform of the real space data using equispaced FFT.

We make a nonuniform mesh such that the width of the smallest cell $\Delta_{min} = 0.0027$, and the width of the largest cell $\Delta_{max} = 0.0054$. Thus the expansion ratio is $\Delta_{max}/\Delta_{min} = 2$.  According to  Gr\"otzbach condition~\cite{Grotzbach:JCP1983}, the mean grid size should be less than $\pi$ times the Kolmogorov and thermal diffusion length scales. For unit $\mathrm{Pr}$, the Kolmogorov and thermal diffusion length scales are equal, and they are estimated using the formula $\eta = L (\mathrm{Pr}^2/(\mathrm{RaNu}))^{1/4} \approx 0.0041$, where $L$ is the box size.  Thus $\pi \eta = 0.013$, hence  $\Delta_{min}$ and $\Delta_{max}$ are less than $\pi \eta$. 

Another important requirement for the DNS is based on the resolution of the thermal boundary layer (BL)~\cite{Grotzbach:JCP1983,Verzicco:JFM2003,Amati:PF2005,Stevens:JFM2010,Shishkina:NJP2010}.  Gr\"otzbach~\cite{Grotzbach:JCP1983} recommends at least two to three points in the BL. Verzicco and Camussi~\cite{Verzicco:JFM2003} and  Amati {\em et al.}~\cite{Amati:PF2005}, however, proposed  more than three grid points  inside the thermal BL. We estimate the width of the boundary layer using the formula $\delta \sim 1/(2\mathrm{Nu})$, in which we keep six points. Thus grid resolution is sufficient for our simulation. We perform grid-independence and $\Delta t$-independence tests of our DNS. The Nusselt numbers computed on  $280^3$ and $300^3$ differ by less than 3\% from the simulation on $256^3$. Similarly, the Nusselt numbers  computed using $ \Delta t = 3 \times 10^{-4}$, $5 \times 10^{-4}$, and $\Delta t  = 10^{-3}$, differ from each other by less than 2\%.

A primary objective of the present paper is to test Taylor's hypothesis.  For the same, we place real-space probes to record time series of the velocity field using which we compute the frequency spectrum  $E(f)$.  To relate our simulations with experiments, we place real space probes near the middle of the six wall, near the middle of the four corner edges, and in the middle of the cube.  We label these probes as front (F), back (B), left (L), right (R), top (T), bottom (Bt), middle corners (MC-I, MC-II, MC-III, MC-IV), and centre (C) respectively.  The number of probes near the wall centres, middle corners, and cubic centres are 5, 3, and 9 respectively.  In Fig.~\ref{fig:probe} we exhibit the probes at B, R, Bt, MC I, MC III, and C.

We record the three components of the velocity field at all the real space probes.  We run our simulation for 80 time units  with constant $\Delta t = 10^{-3}$.  We record the velocity fields at every $10$ steps; thus we have $8 \times 10^3$ data points.  For time-windowed averaging~\cite{Madhow:book}, we break the velocity time-series data into 8 window. Thus each window contains 10 time units, with $10^3$ data points. Then we perform Fourier transform of the velocity components $u_i(t)$ and compute the frequency spectrum $E(f)$ using Eq.~(\ref{eq:Ef}). We report $E(f)$ averaged over 8 time-window for each real space probes. 

In the next section, we will discuss our results based on the  numerical data.  We will focus on the computation of $E(k)$ and $E(f)$.

\section{Results}
\label{sec:results}

\begin{figure}[htbp]
\begin{center}
\includegraphics[scale = 0.75]{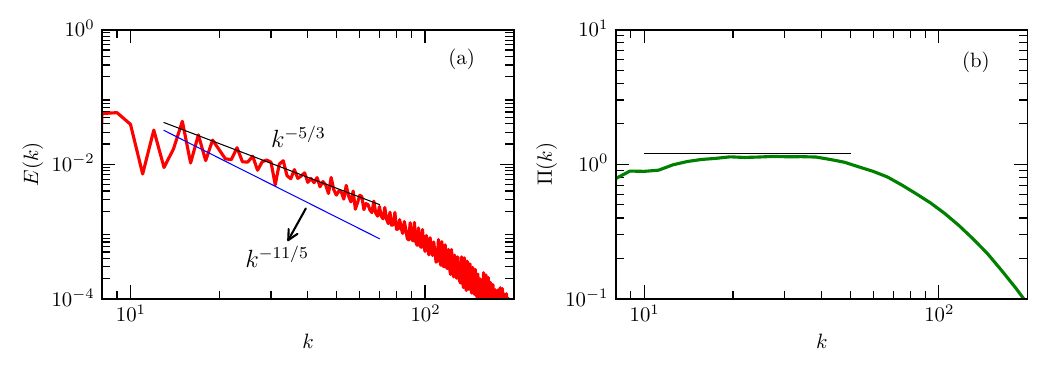}
\end{center}
\setlength{\abovecaptionskip}{0pt}
\caption{For RBC with Prandtl number $\mathrm{Pr} = 1$ and Rayleigh number $\mathrm{Ra} = 10^8$: (a) the kinetic energy spectrum $E(k)$ with $k^{-5/3}$ being a better fit than $k^{-11/5}$; (b) the kinetic energy flux $\Pi(k)$.}
\label{fig:spectrum}
\end{figure}

We interpolate the real space simulation data to a uniform mesh of $256^3$ grids, and then perform Fourier transform using FFT that yields  energy spectrum [$E(k)$] in the wavenumber space. Fig.~\ref{fig:spectrum}(a) demonstrates that the spectrum is Kolmogorov-like, $E(k)=C\epsilon^{2/3}k^{-5/3}$ with $C \approx 1.8$.  We also compute the energy flux using the Fourier modes~\cite{Verma:PR2004}.  The energy flux $\Pi(k)$ plotted in Fig.~\ref{fig:spectrum}(b) shows a constant flux in the inertial range. Thus, our simulation exhibits Kolmogorov's spectrum for RBC, in agreement with the results of Kumar {\em et al.}~\cite{Kumar:PRE2014}, Kumar and Verma~\cite{ Kumar:PRE2015}, and Verma {\em et al.}~\cite{Verma:NJP2017}.

\begin{figure}[htbp]
\begin{center}
\includegraphics[scale = 0.7]{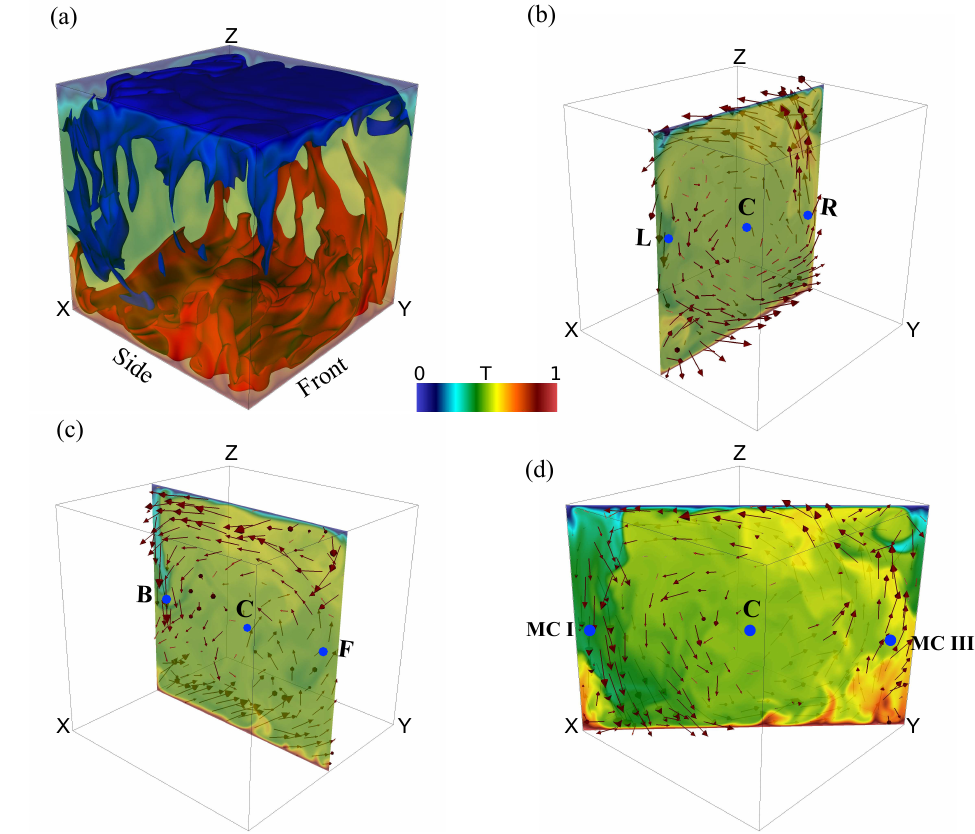}
\end{center}
\setlength{\abovecaptionskip}{0pt}
\caption{For RBC with $\mathrm{Pr} = 1$ and $\mathrm{Ra} = 10^8$: (a)Temperature isosurfaces exhibiting ascending hot plumes (red) and descending cold plume (blue); (b) $xz$ roll with hot  plumes ascending along the right wall and cold plumes descending along the left wall (c) similar $yz$ roll.  (d) Superposition of the two rolls yields diagonal circulation. The corresponding movie for (d) is in the Supplementary Material~\cite{movie}. The color convention of the movie is same as of the present figure.}
\label{fig:flows}
\end{figure}

\setlength{\tabcolsep}{32pt}
\begin{table}[htbp]
\begin{center}
\caption{For RBC with $\mathrm{Pr} = 1$ and $\mathrm{Ra} = 10^8$, the most energetic 5 modes of the flow. $E({\bf k})=|\hat{u}({\bf k})|^2/2$ denotes the modal kinetic energy of the Fourier mode ($k_x$, $k_y$, $k_z$). }
\vspace{2mm}
\begin{tabular}{c  c }  
\hline \hline \\
($k_x$, $k_y$, $k_z$) & $E({\bf k})=|\hat{u}({\bf k})|^2/2$ \\
\\[1mm] \hline 
$(1, 0, 1)$  &$0.126$  \\[1mm]
$(0, 1, 1)$  &$0.050$  \\[1mm]
$(1, 1, 2)$  &$0.003$  \\[1mm]
$(0, 6, 2)$  &$0.002$  \\[1mm]
$(2, 1, 1)$  &$0.002$  \\[1mm]
\hline
\label{table:modes}
\end{tabular}
\end{center}
\end{table}

Thermal plumes and large-scale structures are prominent in thermal convection.  A snapshot of the flow structure in Fig.~\ref{fig:flows}(a) exhibits ascending hot plumes (red) and descending cold plumes (blue). To obtain further details, we analyze the flow velocity at different sections. The three vertical sections exhibited in Fig.~\ref{fig:flows}(b)-(d) clearly demonstrate a {\em large-scale circulation} (LSC)~\cite{Niemela:JFM2001,Sreenivasan:PRE2002,Brown:PRL2005,Xi:PRE2007,Mishra:JFM2011} with  two sets of dominant rolls: in the first roll shown in Fig.~\ref{fig:flows}(b), the hot plumes ascend along the right wall, and the cold plumes descend along the left wall; in the second roll shown in Fig.~\ref{fig:flows}(c), the aforementioned process occurs along the front and back walls.  These two rolls are described by the most energetic velocity Fourier modes ($k_x$, $k_y$, $k_z$) = ($1$,$0$,$1$) and ($0$,$1$,$1$) respectively. The next three most energetic Fourier modes are ($1$,$1$,$2$), ($0$,$6$,$2$), and ($2$,$1$,$1$), but their energies are one order of magnitude lower than those of  ($1$,$0$,$1$) and ($0$,$1$,$1$) modes (see Table~\ref{table:modes}).

Note that the superposition of these modes leads to a strong flow profile in  the diagonal plane shown in Fig.~\ref{fig:flows}(d), but a weak flow profile on the opposite diagonal. The steady LSC is also evident in the movie of  Supplementary Material~\cite{movie}. Note that the movie is from $t=20$ to $t=25$. The presence of LSC suggests that Taylor's hypothesis may be applicable to turbulent convection.   Here the velocity of the mean flow acts as approximate $U_0$.

\begin{figure*}[htbp]
\begin{center}
\includegraphics[scale = 0.8]{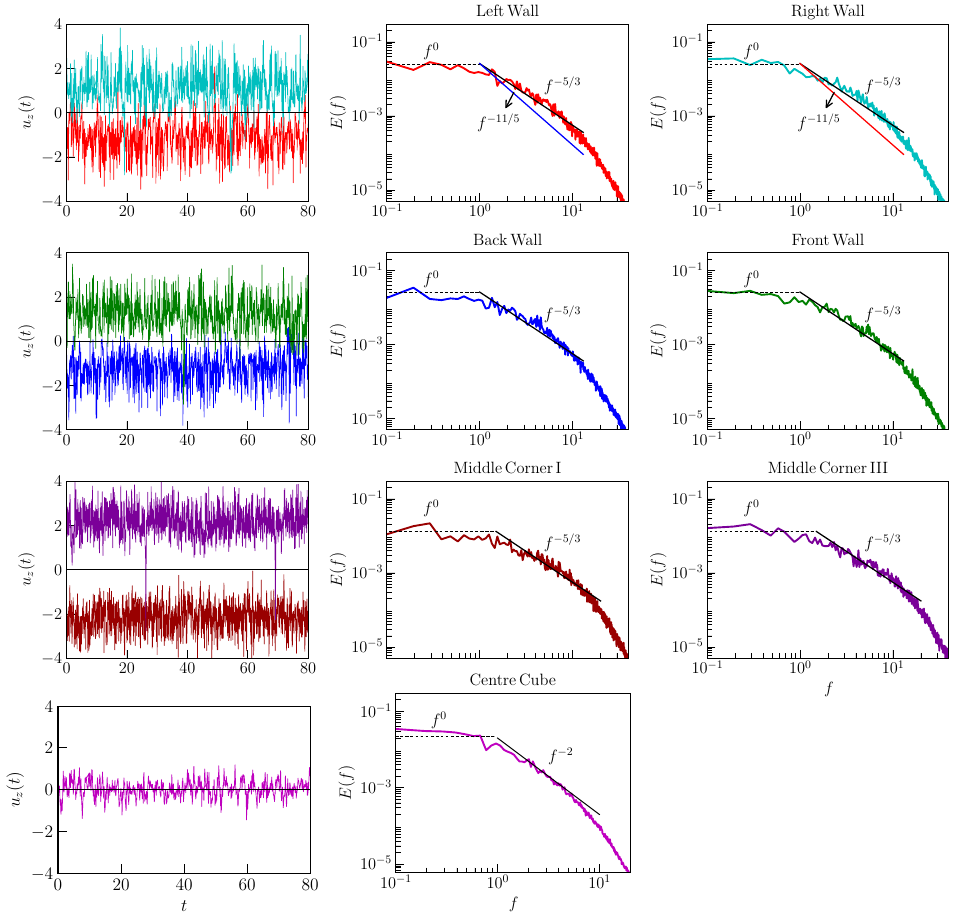}
\end{center}
\setlength{\abovecaptionskip}{0pt}
\caption{For RBC with $\mathrm{Pr} = 1$ and $\mathrm{Ra} = 10^8$, time series $u_z(t)$ measured by the probes, and their corresponding frequency spectra. Left panel: time series for the probes at the left and right walls, back and front walls, middle corners I and III, and centre probes (see Fig.~\ref{fig:probe}). Here time $t$ is in the units of $d/\sqrt{\alpha g \Delta d}$. Middle and right panels: The frequency spectrum $E(f)$ computed for the corresponding probes; $E(f) \sim f^{-5/3}$ fits better than $f^{-11/5}$  for the probes at the side walls and middle corners.   For the centre of the cube, $E(f) \sim f^{-2}$.  At lower frequencies, $E(f) \sim f^{0}$ (white noise).}
\label{fig:probe_spectrum}
\end{figure*}

In the left column of Fig.~\ref{fig:probe_spectrum}, we exhibit the time series
\begin{equation}
u_z(t) = \frac{1}{n} \sum_l u_{z,l}(t),\label{eq:uz}
\end{equation}
measured at L, R, F, B, MC-I, MC-III and C (see Fig.~\ref{fig:probe}). Here $n$ is the number of local probes which are indexed as $l$.  Note that $u_z(t)$  is averaged over all the neighbors, e.g., $u_z(t)$ at B is averaged over the $5$ probes shown in Fig.~\ref{fig:probe}. Here time is in the units of $d/\sqrt{\alpha g \Delta d}$. We observe that $u_z(t)$ of the side walls and corners fluctuate around the  mean values of the LSC.  However, $u_z(t)$ of the centre probes fluctuate around zero, which is due to the absence of any mean velocity at the centre of the cube.

We compute the frequency spectra of the time series as in Eq.~(\ref{eq:Ef}), which are depicted in the middle and right panels of Fig.~\ref{fig:probe_spectrum}  for various set of probes. For the probes at the side walls and mid corners, $E(f) \sim f^{-5/3}$, consistent with the Kolmogorov's phenomenology and Taylor's hypothesis (see middle and right panels of Fig.~\ref{fig:probe_spectrum}).  Here,  the LSC acts as a carrier of the fluctuations. Thus we show that Taylor's hypothesis  can   be  employed  to  the  RBC turbulence  in the presence of a steady LSC.

\begin{figure}[htbp]
\begin{center}
\includegraphics[scale = 0.7]{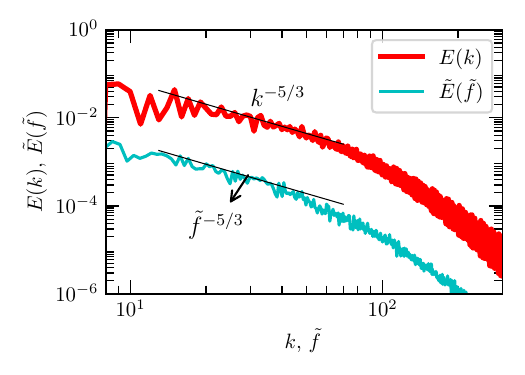} 
\end{center}
\caption{For RBC with $\mathrm{Pr} = 1$ and $\mathrm{Ra} = 10^8$, plot of the wavenumber spectrum $E(k)$ and scaled frequency spectrum of the probe at the left wall (see Fig.~\ref{fig:probe_spectrum}): $f \rightarrow \tilde{f} = f (2\pi)/U_0$ and $E(f) \rightarrow \tilde{E}(\tilde{f}) = E(f) U_0/(2\pi)$. We take $U_0=1$.}
\label{fig:EuEfRBC}
\end{figure}

In Fig.~\ref{fig:EuEfRBC}, we simultaneously plot the wavenumber spectrum and the frequency spectrum at the left wall (see Fig.~\ref{fig:probe_spectrum}) with appropriate scaling---the frequency $f \rightarrow \tilde{f} = f (2\pi)/U_0$ and $E(f) \rightarrow \tilde{E}(\tilde{f}) = E(f) U_0/(2\pi)$. Motivated by the time series $u_z(t)$ of Fig.~\ref{fig:probe_spectrum}, we take $U_0=1$.   We observe that both the spectra exhibit Kolmogorov's spectrum,  but $ \tilde{E}(\tilde{f})$ is several orders of magnitude lower than $E(k)$ in contrast to hydrodynamic turbulence where  $ \tilde{E}(\tilde{f}) \approx E(k)$ (see Fig.~\ref{fig:EuEf_fluid} of Appendix~\ref{appA}). {This is because an LSC roll (one among several LSC rolls) sweeps  fluctuations associated with it. For example, the probes at the left and right walls measure fluctuations advected by the LSC associated with  the Fourier mode ${\bf u}(1,0,1)$; this LSC primarily carries fluctuations in the $xz$ planes whose Fourier modes are of the form $(k_x,0,k_z)$.  On the other hand, for the probes at the back and front walls, the associated LSC would be one corresponding to the Fourier mode ${\bf u}(0,1,1)$ that  primarily advects fluctuations with Fourier modes  $(0,k_y,k_z)$.  Note however $E(k)$ consists of all the fluctuations, be it $(k_x,0,k_z)$ or $(0,k_y,k_z)$.  Hence the frequency spectrum measured by a velocity probe, $E(f)$, is smaller that $E(k)$. Note that  $ \tilde{E}(\tilde{f}) $ of Fig.~\ref{fig:EuEfRBC} is that of only the left wall that corresponds to the mode ${\bf u}(0,1,1)$.}  For a homogeneous and isotropic fluid turbulence, ${\bf U}_0$ advects all forms of random fluctuations, that is,  random fluctuations of arbitrary directions criss-cross the probe during its measurement, thus yielding $ \tilde{E}(\tilde{f}) \approx E(k)$ for hydrodynamic turbulence (see Appendix~\ref{appA}).    { A more refined analysis would clarify this issue.}

There are several RBC experiments  in  rectangular geometry~\cite{Tilgner:PRE1993,Belmonte:PRL1993,Chilla:INCD1993,Maystrenko:PRE2007,Wang:EPJB2003,SUN:JFM2008,Zocchi:PA1990}. Our finding is in agreement with the experimental results of Chill{\`a} {\em et al.}~\cite{Chilla:INCD1993}, which was carried out in a rectangular cell with water as a working fluid. They  showed that  the frequency and wavenumber spectra are approximately equal in the presence of mean flow.

The centre probe, however, exhibits $E(f) \sim f^{-2}$ due to the absence of LSC; this result is same as $E(f) \sim f^{-2}$ observed for the hydrodynamic turbulence with ${\bf U}_0=0$~\cite{Tennekes:book,Landau:Book} (see Appendix~\ref{appA}).   Another interesting feature of $E(f)$ is the robust $f^0$ spectrum (white noise) observed at lower frequencies (see Fig.~\ref{fig:probe_spectrum}).  This feature indicates that the fluctuations at time scales  $t \gtrapprox 1$ (corresponding to $f \lessapprox1$) are uncorrelated.  This behaviour is in sharp contrast to $E(f) \sim f^{-1}$ reported in experiments  exhibiting flow reversals~\cite{Brown:PRL2005,Herault:EPL2015}. The difference is possibly due to the variance of the long-time correlations for reversing and non-reversing velocity signals.

\begin{figure}[htbp]
\begin{center}
\includegraphics[scale = 0.7]{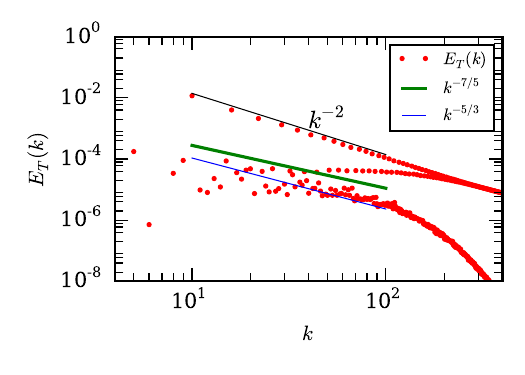} 
\end{center}
\caption{{For RBC simulation with $\mathrm{Pr} = 1$ and $\mathrm{Ra} = 10^8$, plot of the entropy spectrum $E_T(k)$ computed using the temperature field $T({\bf r})$.  The spectrum exhibits dual branch; the upper branch matches with $k^{-2}$ quite well, while the lower branch is fluctuating.}}
\label{fig:ET_K}
\end{figure}

\begin{figure*}[htbp]
\begin{center}
\includegraphics[scale = 0.8]{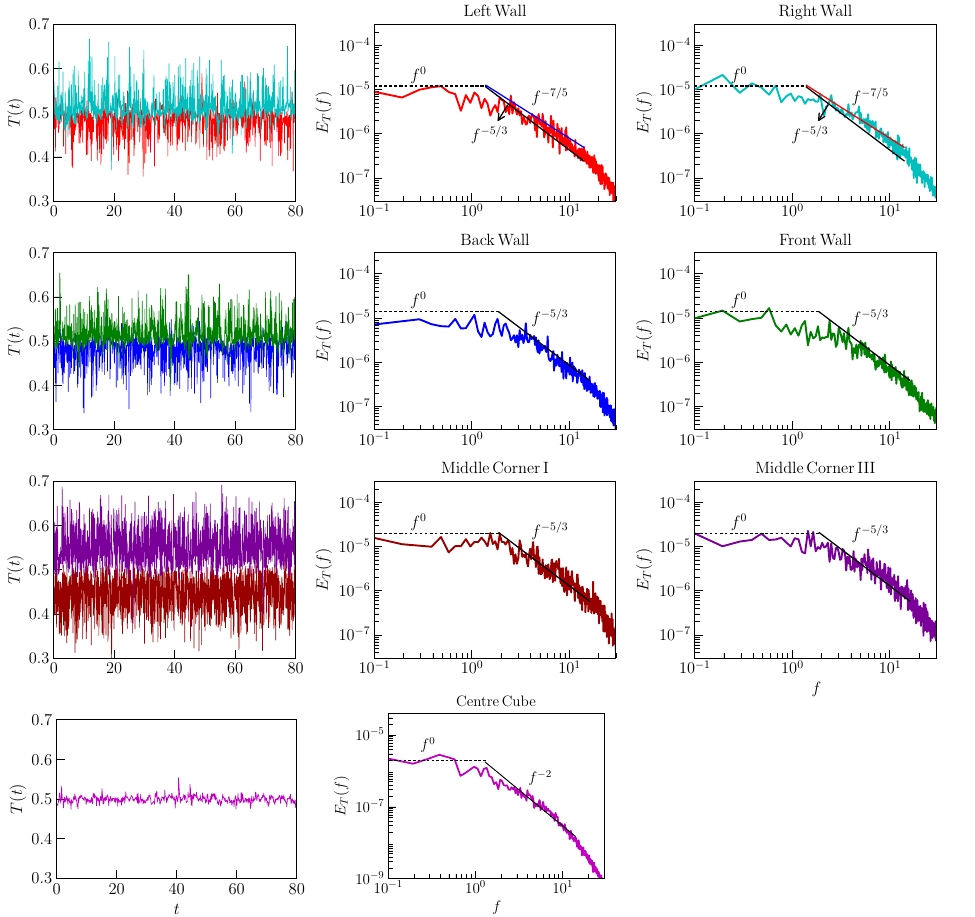}
\end{center}
\caption{{For RBC with $\mathrm{Pr} = 1$ and $\mathrm{Ra} = 10^8$, time series of the temperature, $T(t)$, measured by the probes of Fig.~\ref{fig:probe}.  Middle and right panels: The frequency spectrum of entropy, $E_T(f)$, computed for the corresponding probes.  For  the probes at the side walls and middle corners, $f^{-5/3}$ is a reasonable fit, but the results are somewhat  inconclusive since the two competing exponents $-5/3$ and $-7/5$ are close to each other. For the centre of the cube, $E_T(f) \sim f^{-2}$. At lower frequencies, $E_T(f) \sim f^{0}$.}}
\label{fig:probe_spectrum_T}
\end{figure*}

{ We also compute the entropy spectrum $E_T(k)$ using the  real-space data of the temperature field $T$, as well as frequency spectrum $E_T(f)$ using the time series of the real-space probes.  In Fig.~\ref{fig:ET_K}, we plot the entropy spectrum $E_T(k)$ that shows dual branch with the upper branch scaling as $k^{-2}$.  Mishra and Verma~\cite{Mishra:PRE2010},  Pandey {\em et al.}~\cite{Pandey:PRE2016}, and Verma {\em et al.}~\cite{Verma:NJP2017} showed the dominant temperature modes $T(k_x=0,k_y=0,k_z=2n)$, where $n$ is an integer, constitute the $k^{-2}$ branch of $E_T(k)$, for which the modes $T(0,0,2n)$ play a critical role. 

In Fig.~\ref{fig:probe_spectrum_T}, we plot the time series of the temperature field measured at various probe locations, and their corresponding entropy spectra. For the probes at the side walls and middle corners, the Kolmogorov--Obukhov spectrum ($f^{-5/3}$) appears to fit better than  Bolgiano--Obukhov spectrum ($f^{-7/5}$), but the fits are not very conclusive.  One reason for the ambiguity is that the exponents $-5/3$ and $-7/5$ are quite close to each other.   We observe that the frequency spectrum of the velocity field is  more conclusive than  that of the temperature field, though a more detailed study in this direction is required. }

\section{Conclusions}
\label{sec:conc}

 { A primary objective of this paper is to test Taylor's hypothesis for turbulent convection in  a cube.  To this end, }we performed the direct numerical simulation of Rayleigh-B\'{e}nard convection in a closed cubical box for $\mathrm{Pr}=1$ and $\mathrm{Ra}=10^8$ and studied the energy spectrum using the numerical data in space-domain and time-domain. We placed the real space probes in the simulation box and measured the time series of the velocity field.   For the velocity field, the wavenumber energy spectrum as well as the frequency spectrum exhibit Kolmogorov's spectrum.  We observe that the kinetic energy flux is constant.  These observations demonstrate that RBC has a similar scaling as of hydrodynamic turbulence, rather than Bolgiano--Obukhov's  scaling.  These results are consistent with recent works~\cite{Kumar:PRE2014, Verma:NJP2017}.

{ In our numerical simulation we observed that  $E(k) \sim k^{-5/3}$ and $E(f)\ \sim f^{-5/3}$, hence we conclude that Taylor's hypothesis is applicable in a cube for the parameters employed in this paper.} The analysis of the flow structures of  RBC and their associated Fourier modes demonstrate  presence of a steady large-scale circulation  in the flow.  Such a mean flow enables an application of Taylor's hypothesis to turbulent convection, which is why both $E(k)$ and $E(f)$ show Kolmogorov's spectrum.  

{ Note that turbulent convection in a cylinder exhibits azimuthal rotation or  reversals of  LSC that may make  application of Taylor's hypothesis questionable. {The ambiguities in the spectral exponents in earlier experimental results~\cite{Wu:PRL1990,Cioni:EPL1995,Zhou:PRL2001,Skrbek:PRE2002,Niemela:NATURE2000,Shang:PRE2001,Sun:PRL2006} are probably due to unsteady nature of LSC. For these reasons, we advocate usage of rectangular rather than a cylindrical or spherical geometry for spectral studies in thermal convection because LSC is more steady  in a box compared to a cylinder.  We, however, remark the recent thermal convection simulations in a cube~\cite{Foroozani:PRE2017,Vasiliev:IJHMT2016}  exhibits flow reversals for a set of parameters.  Hence, we need to compute $E(k)$ and $E(f)$ for such systems to ascertain the applicability of Taylor's hypothesis in a cube.}}

He {\em et al.}~\cite{He:PRE2010} measured the temperature field at various real-space probes, and then computed the frequency spectrum of  temperature field by invoking elliptic approximation and deduced that $E_T(k) \sim k^{-1.35}$.  This spectrum does not match with the spectrum reported by Kumar {\em et al.}~\cite{Kumar:PRE2014} and Verma {\em et al.}~\cite{Verma:NJP2017}. The discrepancy is possibly due to fact that temperature field exhibits dual branch, which is not captured by the frequency spectrum of temperature. Thus the $E(f)$ of the velocity field reported in this paper is a more concrete demonstration of the energy spectrum of turbulent convection. {  We remark that the competing spectral exponents of the temperature field, $-5/3$ and$-7/5$, are too close for a conclusive contrast. The corresponding exponents for the velocity field are $-5/3$ and $-11/5$, which are relatively further apart. }   These results suggest that the velocity field or the velocity field extrapolated from the temperature measurement would provide better handle on the energy spectrum than using temperature probe. 

In summary, our numerical simulation of RBC in a cube demonstrates Kolmogorov's spectrum for both wavenumber and frequency spectra.  The correspondence between the two spectra is due to the {\em steady large-scale circulation} and Taylor's hypothesis.  {  We, however, caution that more work is required for reaching a definite conclusion.}

\appendix
\section{Taylor's hypothesis in hydrodynamic turbulence}
\label{appA}
The dynamical equations for the incompressible velocity field are
 \begin{eqnarray}
\frac{\partial \bf u}{\partial t} + (\bf U_0 \cdot \nabla) \bf u + (\bf u \cdot \nabla) \bf u & = & - \frac{1}{\rho_0} \nabla p+ \nu \nabla^2 \bf u  + {\bf f} , \label{eq:u_fluid} \\
\nabla \cdot \bf u & = & 0 \label{eq:inc_fluid},
\end{eqnarray}
where ${\bf u}, p, {\bf f}$ are the velocity, pressure, and external force fields respectively, and $\nu$ is the kinematic viscosity.  To test Taylor's hypothesis for hydrodynamic turbulence, we numerically solve the above equations with  ${\bf U}_0 = 0$ in a periodic box of dimension $(2 \pi)^3$  using a pseudospectral code Tarang~\cite{Chatterjee:JPDC17}. In the simulation, we employed a   random forcing~\cite{Reddy:PF2014} to the flow in the wavenumber band $1 \leq k \leq 3$ on  $512^3$ grid. We also use a fourth-order Runge-Kutta  scheme for time stepping and 2/3 rule for aliasing.  We continue our simulation till it reaches a steady state. \begin{figure}[htbp]
\begin{center}
\includegraphics[scale = 0.65]{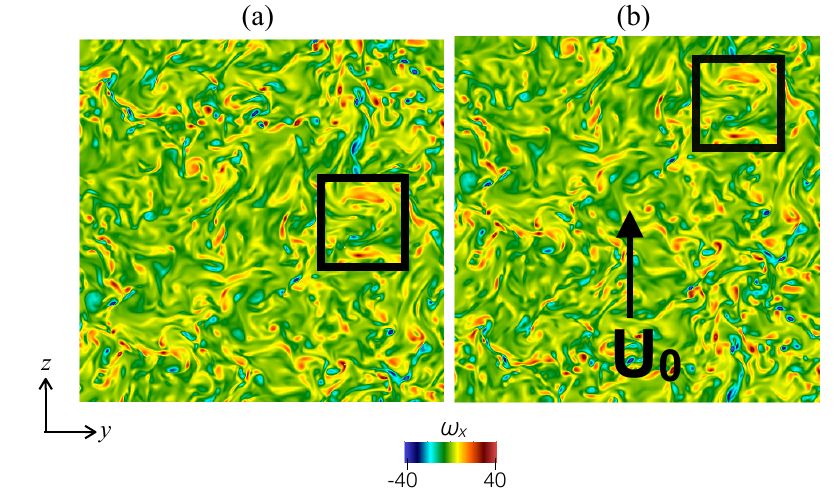}
\end{center}
\setlength{\abovecaptionskip}{0pt}
\caption{For hydrodynamic turbulence simulation with $\mathrm{Re} \approx1100$, and ${\bf U}_0 = 0$ and ${\bf U}_0 = 10 \hat{z}$: The density plots of the vorticity component $\omega_x$ of a vertical cross-section for (a) ${\bf U}_0 = 0$ and (b) ${\bf U}_0 =10 \hat{z}$.  The flow in (b) is shifted upward by ${\bf U}_0\tau$ compared to (a).}
\label{fig:density_plot}
\end{figure} Once a steady state is reached, we initiate two new runs with  ${\bf U}_0 = 0$ and ${\bf U}_0 = 10 \hat{z}$ using the above final state as the initial condition.  We carry out the two simulations till $t=1$, where time units is $2\pi/u_{\mathrm{rms}}$. Here $u_{\mathrm{rms}}$ is the rms velocity of the flow, computed as the volume average of the magnitude of the velocity field.  

\begin{figure}[htbp]
\begin{center}
\includegraphics[scale = 0.7]{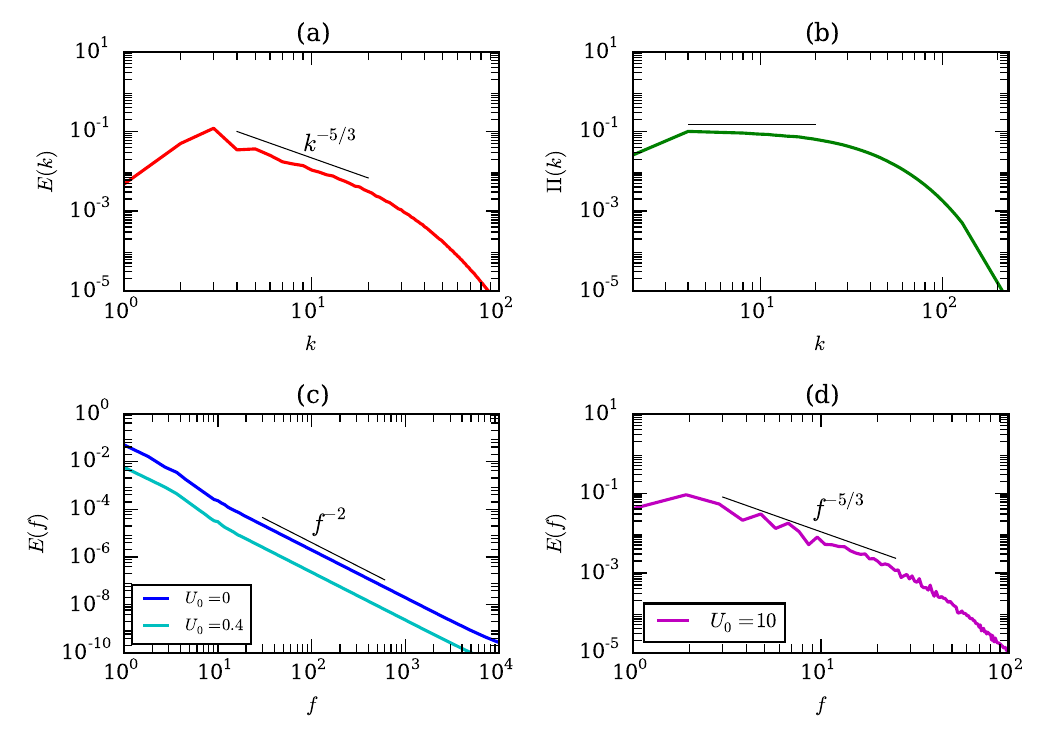}
\end{center}
\setlength{\abovecaptionskip}{0pt}
\caption{For hydrodynamic turbulence simulation with $\mathrm{Re} \approx1100$, and ${\bf U}_0 = 0$ and ${\bf U}_0 = 10 \hat{z}$:  (a) energy spectra and (b) energy fluxes are the same for both the flows; they follow Kolmogorov's model. (c) Frequency spectra $E(f) \sim f^{-2}$ for ${\bf U}_0 = 0$ and $0.4$; (d) For ${\bf U}_0 = 10$, $E(f) \sim f^{-5/3}$ consistent with Taylor's hypothesis.}
\label{fig:spectrum_flux}
\end{figure}

\begin{figure}[htbp]
\begin{center}
\includegraphics[scale = 0.8]{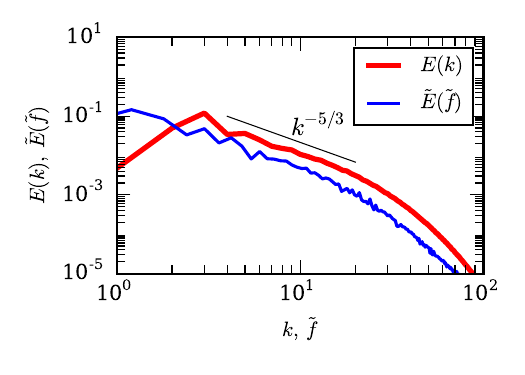}
\end{center}
\setlength{\abovecaptionskip}{0pt}
\caption{For hydrodynamic turbulence simulation with  ${\bf U}_0 = 10 \hat{z}$, plot of the wavenumber spectrum $E(k)$ and scaled frequency spectrum for the real space probes: $f \rightarrow \tilde{f} = f (2\pi)/U_0$ and $E(f) \rightarrow \tilde{E}(\tilde{f}) = E(f) U_0/(2\pi)$.}
\label{fig:EuEf_fluid}
\end{figure}

The Reynolds number of the flows $\mathrm{Re}= UL/\nu \approx 1100$, where $L$, $U$ are respectively the length and velocity scales of the flow.  In Fig.~\ref{fig:density_plot}, we illustrate the vorticity component $\omega_x$ at the same cross-section and the same time in the two boxes. Figure~\ref{fig:density_plot}(b) is an upward translation by ${\bf U}_0 \tau$, where $\tau$ is the time interval, of Fig.~\ref{fig:density_plot}(a), thus indicating a vertical motion of the flow due to ${\bf U}_0$.  We compute the energy spectrum $E(k)$ and the energy flux $\Pi(k)$ for both the datasets. As expected, these quantities are identical for both the boxes, and they are plotted in Fig.~\ref{fig:spectrum_flux}(a,b) respectively. In the inertial range, we observe Kolmogorov's spectrum.

We record the time series of the velocity field at $50$ random locations, and then compute their frequency spectra $E(f)$. Figs.~\ref{fig:spectrum_flux}(c,d) exhibit the averaged $E(f)$ computed using the time series recorded by $50$ randomly-located real-space probes for ${\bf U}_0 = 0$ and ${\bf U}_0 = 10 \hat{z}$ respectively.  For ${\bf U}_0 = 10 \hat{z}$, $E(f) \sim f^{-5/3}$, in accordance with Taylor's hypothesis~\cite{Taylor:PRSLA1938,Landau:Book} since $2\pi f = U_0k$. However, for homogeneous and isotropic turbulence with ${\bf U}_0 = 0$, we obtain $E(f) \sim f^{-2}$ since $f \sim \epsilon^{1/3}k^{2/3}$ from Kolmogorov's theory~\cite{Tennekes:book}. Interestingly, we also observe   $E(f) \sim f^{-2}$ for small $ U_0$ when $\epsilon^{1/3}k^{2/3} > U_0 k$ (see Fig.~\ref{fig:spectrum_flux}(c) for ${\bf U}_0 = 0.4 \hat{z}$).

In Fig.~\ref{fig:EuEf_fluid} we  plot the wavenumber spectrum and scaled frequency spectrum:$f \rightarrow \tilde{f} = f (2\pi)/U_0$ and $E(f) \rightarrow \tilde{E}(\tilde{f}) = E(f) U_0/(2\pi)$, where $U_0=10$.   We observe that both the spectra exhibit Kolmogorov's spectrum,  and $ \tilde{E}(\tilde{f}) \approx E(k)$.  Thus we demonstrate consistency with the Taylor's hypothesis for hydrodynamic turbulence.

\pagebreak
\section*{Acknowledgements}
This work was supported by a research grants (Grant No. SERB/F/3279) from Science and Engineering Research Board, India and (Grant No. PLANEX/PHY/2015239) from Indian Space Research Organisation, India.We thank Sagar Chakraborty for valuable suggestions. Our numerical simulations were performed on {\em HPC} and {\em Chaos}  clusters of IIT Kanpur.



\begin{thebibliography}{99}

\bibitem{Chandrasekhar:book}
{Chandrasekhar} S. 1961 {\em Hydrodynamic and Hydromagnetic Stability}.
New York: Dover Publications.

\bibitem{JKB:book}
{Bhattacharjee} JK. 1987 {\em Convection and Chaos in Fluids}.
Singapore: World Scientific.

\bibitem{Bodenschatz:ARFM2000}
{Bodenschatz} E, {Pesch} W, {Ahlers} G. 2000  Recent developments in
  Rayleigh-B{\'e}nard convection. {\em Ann. Rev. Fluid Mech.} \textbf{32},
  709--778.

\bibitem{Ahlers:RMP2009}
{Ahlers} G, {Grossmann} S, {Lohse} D. 2009  Heat transfer and large scale
  dynamics in turbulent Rayleigh-B{\'e}nard convection. {\em Rev. Mod. Phys.}
  \textbf{81}, 503--537.

\bibitem{Lohse:ARFM2010}
{Lohse} D, {Xia} KQ. 2010  Small-scale properties of turbulent
  Rayleigh-B{\'e}nard convection. {\em Ann. Rev. Fluid Mech.} \textbf{42},
  335--364.

\bibitem{Kolmogorov:DANS1941a}
{Kolmogorov} AN. 1941  The local structure of turbulence in incompressible
  viscous fluid for very large Reynolds numbers. {\em Dokl. Akad. Nauk SSSR}
  \textbf{30}, 9--13.

\bibitem{Bolgiano:JGR1959}
{Bolgiano} R. 1959  Turbulent spectra in a stably stratified atmosphere. {\em
  J. Geophys. Res.} \textbf{64}, 2226.

\bibitem{Obukhov:DANS1959}
{Obukhov} AN. 1959  Effect of Archimedean forces on the structure of the
  temperature field in a turbulent flows. {\em Dokl. Akad. Nauk SSSR}
  \textbf{125}, 1246.

\bibitem{Procaccia:PRL1989}
{Procaccia} I, {Zeitak} R. 1989  Scaling exponents in nonisotropic convective
  turbulence. {\em Phys. Rev. Lett.} \textbf{62}, 2128--2131.

\bibitem{Lvov:PD1992}
{L'vov} VS, {Falkovich} GE. 1992  Conservation laws and two-flux spectra of
  hydrodynamic convective turbulence. {\em Physica D} \textbf{57}, 85.

\bibitem{Kumar:PRE2014}
Kumar A, Chatterjee AG, Verma MK. 2014  Energy spectrum of buoyancy-driven
  turbulence. {\em Phys. Rev. E} \textbf{90}, 023016.

\bibitem{Verma:NJP2017}
Verma MK, Kumar A, Pandey A. 2017  {Phenomenology of buoyancy-driven
  turbulence: recent results}. {\em New J. Phys.} \textbf{19}, 025012.

\bibitem{Kumar:PRE2015}
Kumar A, Verma MK. 2015  {Shell model for buoyancy-driven turbulence}. {\em
  Phys. Rev. E} \textbf{91}, 043014.

\bibitem{Mishra:PRE2010}
{Mishra} PK, {Verma} MK. 2010  Energy spectra and fluxes for
  Rayleigh-B{\'e}nard convection. {\em Phys. Rev. E} \textbf{81}, 056316.

\bibitem{Sun:PRL2006}
{Sun} C, {Zhou} Q, {Xia} KQ. 2006  Cascades of velocity and temperature
  fluctuations in buoyancy-driven thermal turbulence. {\em Phys. Rev. Lett.}
  \textbf{97}, 144504.

\bibitem{Zhou:JFM2008}
Zhou Q, Sun C, Xia KQ. 2008  Experimental investigation of homogeneity,
  isotropy, and circulation of the velocity field in buoyancy-driven
  turbulence. {\em J. Fluid Mech.} \textbf{598}, 361--372.

\bibitem{Kunnen:PRE2008}
{Kunnen} R, {Clercx} H, {Geurts} B, {Bokhoven} LV, {Akkermans} R, {Verzicco} R.
  2008  Numerical and experimental investigation of structure-function scaling
  in turbulent Rayleigh-B{\'e}nard convection. {\em Phys. Rev. E} \textbf{77},
  016302.

\bibitem{Zhou:JFM2011}
Zhou Q, Li CM, Lu ZM, Liu YL. 2011  Experimental investigation of longitudinal
  space--time correlations of the velocity field in turbulent
  Rayleigh--B{\'e}nard convection. {\em J. Fluid Mech.} \textbf{683}, 94--111.

\bibitem{Wu:PRL1990}
{Wu} XZ, {Kadanoff} L, {Libchaber} A, {Sano} M. 1990  Frequency power spectrum
  of temperature fluctuations in free convection. {\em Phys. Rev. Lett.}
  \textbf{64}, 2140--2143.

\bibitem{Chilla:INCD1993}
{Chill\`a} F, {Ciliberto} S, {Innocenti} C, {Pampaloni} E. 1993  Boundary layer
  and scaling properties in turbulent thermal convection. {\em Nuovo Cimento D}
  \textbf{15}, 1229.

\bibitem{Cioni:EPL1995}
{Cioni} S, {Ciliberto} S, {Sommeria} J. 1995  Temperature structure functions
  in turbulent convection at low Prandtl number. {\em EPL} \textbf{32},
  413--418.

\bibitem{Zhou:PRL2001}
{Zhou} SQ, {Xia} KQ. 2001  Scaling properties of the temperature field in
  convective turbulence. {\em Phys. Rev. Lett.} \textbf{87}, 064501.

\bibitem{Skrbek:PRE2002}
{Skrbek} L, {Niemela} JJ, {Sreenivasan} KR, {Donnelly} RJ. 2002  Temperature
  structure functions in the Bolgiano regime of thermal convection. {\em Phys.
  Rev. E} \textbf{66}, 36303.

\bibitem{Niemela:NATURE2000}
{Niemela} JJ, {Skrbek} L, {Sreenivasan} KR, {Donnelly} RJ. 2000  Turbulent
  convection at very high Rayleigh numbers. {\em Nature} \textbf{404},
  837--840.

\bibitem{Shang:PRE2001}
{Shang} XD, {Xia} KQ. 2001  Scaling of the velocity power spectra in turbulent
  thermal convection. {\em Phys. Rev. E} \textbf{64}, 065301.

\bibitem{Taylor:PRSLA1938}
{Taylor} GI. 1938  The spectrum of turbulence. {\em Proc. R. Soc. Lond. A}
  \textbf{164}, 476.

\bibitem{Tennekes:book}
{Tennekes} H, {Lumley} JL. 1972 {\em A First Course in Turbulence}.
Cambridge, Massachusetts: The MIT Press.

\bibitem{Landau:Book}
{Landau} LD, {Lifshitz} EM. 1987 {\em Fluid Mechanics}.
Oxford: Pergamon Press.

\bibitem{He:PRE2006}
He GW, Zhang JB. 2006  Elliptic model for space-time correlations in turbulent
  shear flows. {\em Phys. Rev. E} \textbf{73}, 055303.

\bibitem{He:Acta2014}
He X, Tong P. 2014  {Space-time correlations in turbulent Rayleigh-B{\'e}nard
  convection}. {\em Acta Mech. Sinica} \textbf{30}, 457--467.

\bibitem{He:ARFM2016}
He G, Jin G, Yang Y. 2017  {Space-time correlations and dynamic coupling in
  turbulent flows}. {\em Annu. Rev. Fluid Mech.} \textbf{49}, 51--70.

\bibitem{Cholemari:JFM2006}
Cholemari MR, Arakeri JH. 2006  {A model relating Eulerian spatial and temporal
  velocity correlations}. {\em J. Fluid Mech.} \textbf{551}, 19--29.

\bibitem{He:PRE2010}
He X, He G, Tong P. 2010  Small-scale turbulent fluctuations beyond Taylor's
  frozen-flow hypothesis. {\em Phys. Rev. E} \textbf{81}, 065303.

\bibitem{He:PRE2011}
He X, Tong P. 2011  {Kraichnan{\textquoteright}s random sweeping hypothesis in
  homogeneous turbulent convection}. {\em Phys. Rev. E} \textbf{83}, 037302.

\bibitem{Castaing:PRL1990}
{Castaing} B. 1990  Scaling of turbulent spectra. {\em Phys. Rev. Lett.}
  \textbf{65}, 3209.

\bibitem{Mashiko:PRE2004}
{Mashiko} T, {Tsuji} Y, {Mizuno} T, {Sano} M. 2004  Instantaneous measurement
  of velocity fields in developed thermal turbulence in mercury. {\em Phys.
  Rev. E} \textbf{69}, 036306.

\bibitem{Ashkenazi:PRL1999b}
{Ashkenazi} S, {Steinberg} V. 1999  Spectra and statistics of velocity and
  temperature fluctuations in turbulent convection. {\em Phys. Rev. Lett.}
  \textbf{83}, 4760.

\bibitem{Bershadskii:PRE2004}
{Bershadskii} A, {Niemela} JJ, {Praskovsky} A, {Sreenivasan} KR. 2004
  ``Clusterization'' and intermittency of temperature fluctuations in turbulent
  convection. {\em Phys. Rev. E} \textbf{69}, 56314.

\bibitem{Arakeri:CS2000}
Arakeri JH, Avila FE, Dada JM, Tovar RO. 2000  Convection in a long vertical
  tube due to unstable stratification - a new type of turbulent flow?. {\em
  Curr. Sci.} \textbf{79}, 859--866.

\bibitem{Pawar:POF2016}
Pawar SS, Arakeri JH. 2016  {Kinetic energy and scalar spectra in high Rayleigh
  number axially homogeneous buoyancy driven turbulence}. {\em Phys. Fluids}
  \textbf{28}, 065103--8.

\bibitem{Grossmann:PRL1991}
{Grossmann} S, {Lohse} D. 1991  Fourier-Weierstrass mode analysis for thermally
  driven turbulence. {\em Phys. Rev. Lett.} \textbf{67}, 445--448.

\bibitem{Borue:JSC1997}
{Borue} V, {Orszag} SA. 1997  Turbulent convection driven by a constant
  temperature gradient. {\em J. Sci. Comput.} \textbf{12}, 305--351.

\bibitem{Skandera:HPCISEG2SBH2009}
{\v{S}kandera} D, {Busse} A, M{\"u}ller WC. 2008  Scaling Properties of
  Convective Turbulence. {\em High Performance Computing in Science and
  Engineering,Workshop (Springer, Berlin)} p. 387.

\bibitem{Rincon:JFM2006}
{Rincon} F. 2006  Anisotropy, inhomogeneity and inertial-range scalings in
  turbulent convection. {\em J. Fluid Mech.} \textbf{563}, 43.

\bibitem{Verzicco:JFM2003}
{Verzicco} R, {Camussi} R. 2003  Numerical experiments on strongly turbulent
  thermal convection in a slender cylindrical cell. {\em J. Fluid Mech.}
  \textbf{477}, 19--49.

\bibitem{Camussi:EJMF2004}
{Camussi} R, {Verzicco} R. 2004  Temporal statistics in high Rayleigh number
  convective turbulence. {\em Eur. J. of Mech. /B Fluids} \textbf{23}, 427.

\bibitem{Calzavarini:PRE2002}
{Calzavarini} E, {Toschi} F, {Tripiccione} R. 2002  Evidences of
  Bolgiano-Obhukhov scaling in three-dimensional Rayleigh-B{\'e}nard
  convection. {\em Phys. Rev. E} \textbf{66}, 016304.

\bibitem{De:IJHFF2017}
De AK, Eswaran V, Mishra PK. 2017  {Scalings of heat transport and energy
  spectra of turbulent Rayleigh-B{\'e}nard convection in a large-aspect-ratio
  box}. {\em Int. J. Heat Fluid Flow} \textbf{67}, 111--124.

\bibitem{Kaczorowski:JFM2013}
Kaczorowski M, Xia KQ. 2013  {Turbulent flow in the bulk of
  Rayleigh{\textendash}B{\'e}nard convection: Small-scale properties in a cubic
  cell}. {\em J. Fluid Mech.} \textbf{722}, 596--617.

\bibitem{Kerr:JFM1996}
{Kerr} RM. 1996  Rayleigh number scaling in numerical convection. {\em J. Fluid
  Mech.} \textbf{310}, 139--179.

\bibitem{Nath:up2016}
{Nath} D, {Pandey} A, {Kumar} A, {Verma} MK. 2016  Near isotropic behaviour of
  turbulent thermal convection. {\em Phys. Rev. Fluids} \textbf{1}, 064302.

\bibitem{Niemela:JFM2001}
{Niemela} JJ, {Skrbek} L, {Sreenivasan} KR, {Donnelly} RJ. 2001  The wind in
  confined thermal convection. {\em J. Fluid Mech.} \textbf{449}, 169.

\bibitem{Sreenivasan:PRE2002}
{Sreenivasan} KR, {Bershadskii} A, {Niemela} JJ. 2002  Mean wind and its
  reversal in thermal convection. {\em Phys. Rev. E} \textbf{65}, 056306.

\bibitem{Brown:PRL2005}
{Brown} E, {Nikolaenko} A, {Ahlers} G. 2005  Reorientation of the large-scale
  circulation in turbulent Rayleigh-B{\'e}nard convection. {\em Phys. Rev.
  Lett.} \textbf{95}, 084503.

\bibitem{Xi:PRE2007}
{Xi} HD, {Xia} KQ. 2007  Cessations and reversals of the large-scale
  circulation in turbulent thermal convection. {\em Phys. Rev. E} \textbf{75},
  066307.

\bibitem{Mishra:JFM2011}
{Mishra} PK, {De} AK, {Verma} MK, {Eswaran} V. 2011  Dynamics of reorientations
  and reversals of large-scale flow in Rayleigh-B{\'e}nard convection. {\em J.
  Fluid Mech.} \textbf{668}, 480--499.

\bibitem{Foroozani:PRE2017}
Foroozani N, Niemela JJ, Armenio V, Sreenivasan KR. 2017  {Reorientations of
  the large-scale flow in turbulent convection in a cube}. {\em Phys. Rev. E}
  \textbf{95}, 033107.

\bibitem{Vasiliev:IJHMT2016}
Vasiliev A, Sukhanovskii A, Frick P, Budnikov A, Fomichev V, Bolshukhin M,
  Romanov R. 2016  High Rayleigh number convection in a cubic cell with
  adiabatic sidewalls. {\em Int. J. Heat Mass Transfer} \textbf{102}, 201 --
  212.

\bibitem{Madhow:book}
Madhow U. 2008 {\em Fundamentals of Digital Communication}.
Cambridge: Cambridge University Press.

\bibitem{Dar:PD2001}
{Dar} G, {Verma} MK, {Eswaran} V. 2001  Energy transfer in two-dimensional
  magnetohydrodynamic turbulence: Formalism and numerical results. {\em Physica
  D} \textbf{157}, 207--225.

\bibitem{Verma:PR2004}
{Verma} MK. 2004  Statistical theory of magnetohydrodynamic turbulence: Recent
  results. {\em Phys. Rep.} \textbf{401}, 229--380.

\bibitem{OpenFOAM}
OpenFOAM. 2015 {\em The open source CFD toolbox, www.openfoam.org}.

\bibitem{Grotzbach:JCP1983}
{Gr\"otzbach} G. 1983  Saptial resolution requirement for direct numerical
  simulation of Rayleigh-B\'{e}nard convection. {\em J. Comp. Phys.}
  \textbf{49}, 241--264.

\bibitem{Amati:PF2005}
{Amati} G, {Koal} K, {Massaioli} F, {Sreenivasan} KR, {Verzicco} R. 2005
  Turbulent thermal convection at high Rayleigh numbers for a Boussinesq fluid
  of constant Prandtl number. {\em Phys. Fluids} \textbf{17}, 1701.

\bibitem{Stevens:JFM2010}
{Stevens} R, {Verzicco} R, {Lohse} D. 2010  Radial boundary layer structure and
  Nusselt number in Rayleigh-B{\'e}nard convection. {\em J. Fluid Mech.}
  \textbf{643}, 495--507.

\bibitem{Shishkina:NJP2010}
{Shishkina} O, {Stevens} R, {Grossmann} S, {Lohse} D. 2010  Boundary layer
  structure in turbulent thermal convection and its consequences for the
  required numerical resolution. {\em New J. Phys.} \textbf{12}, 075022.

\bibitem{movie}
 See supplementary material for a movie of the convective flow at a vertical
  section.

\bibitem{Tilgner:PRE1993}
Tilgner A, Belmonte A, Libchaber A. 1993  Temperature and velocity profiles of
  turbulent convection in water. {\em Phys. Rev. E} \textbf{47}, R2253--R2256.

\bibitem{Belmonte:PRL1993}
Belmonte A, Tilgner A, Libchaber A. 1993  Boundary layer length scales in
  thermal turbulence. {\em Phys. Rev. Lett.} \textbf{70}, 4067--4070.

\bibitem{Maystrenko:PRE2007}
Maystrenko A, Resagk C, Thess A. 2007  Structure of the thermal boundary layer
  for turbulent Rayleigh-B\'enard convection of air in a long rectangular
  enclosure. {\em Phys. Rev. E} \textbf{75}, 066303.

\bibitem{Wang:EPJB2003}
Wang J, Xia KQ. 2003  {Spatial variations of the mean and statistical
  quantities in the thermal boundary layers of turbulent convection}. {\em Eur.
  Phys. J. B} \textbf{32}, 127--136.

\bibitem{SUN:JFM2008}
Sun C, Cheung YH, Xia KQ. 2008  {Experimental studies of the viscous boundary
  layer properties in turbulent Rayleigh{\textendash}B{\'e}nard convection}.
  {\em J. Fluid Mech.} \textbf{605}, 1--35.

\bibitem{Zocchi:PA1990}
Zocchi G, Moses E, Libchaber A. 1990  {Coherent structures in turbulent
  convection, an experimental study}. {\em Physica A} \textbf{166}, 387--407.

\bibitem{Herault:EPL2015}
Herault J, P\'{e}tr\'{e}lis F, Fauve S. 2015  Experimental observation of $1/f$
  noise in quasi-bidimensional turbulent flows. {\em EPL} \textbf{111}, 44002.

\bibitem{Pandey:PRE2016}
Pandey A, Kumar A, Chatterjee AG, Verma MK. 2016  {Dynamics of large-scale
  quantities in Rayleigh-B{\'e}nard convection}. {\em Phys. Rev. E}
  \textbf{94}, 053106--7.

\bibitem{Chatterjee:JPDC17}
Chatterjee AG, Verma MK, Kumar A, Samtaney R, Hadri B, Khurram R. 2017
  {Scaling of a Fast Fourier Transform and a pseudo-spectral fluid solver up to
  196608 cores}. {\em J. Parallel Distrib. Comput.} \textbf{113}, 77--91.

\bibitem{Reddy:PF2014}
{Reddy} KS, {Verma} MK. 2014  Strong anisotropy in quasi-static
  magnetohydrodynamic turbulence for high interaction parameters. {\em Phys.
  Fluids} \textbf{26}, 025109.

\end{thebibliography}

\end{document}